\documentstyle{mn}
\input epsf.sty
\newif\ifAMStwofonts
\AMStwofontstrue
 
\ifoldfss
  \ifCUPmtlplainloaded \else
    \NewTextAlphabet{textbfit} {cmbxti10} {}
    \NewTextAlphabet{textbfss} {cmssbx10} {}
    \NewMathAlphabet{mathbfit} {cmbxti10} {} 
    \NewMathAlphabet{mathbfss} {cmssbx10} {} 
  \fi
  \ifAMStwofonts
    \ifCUPmtlplainloaded \else
      \NewSymbolFont{upmath} {eurm10}
      \NewSymbolFont{AMSa} {msam10}
      \NewMathSymbol{\upi}     {0}{upmath}{19}
      \NewMathSymbol{\umu}     {0}{upmath}{16}
      \NewMathSymbol{\upartial}{0}{upmath}{40}
      \NewMathSymbol{\leqslant}{3}{AMSa}{36}
      \NewMathSymbol{\geqslant}{3}{AMSa}{3E}

    \fi
  \fi
\fi 
 
\ifnfssone
  \newmathalphabet{\mathit}
  \addtoversion{normal}{\mahit}{cmr}{m}{it}
  \addtoversion{bold}{\mathit}{cmr}{bx}{it}
  \newmathalphabet{\mathbfit} 
  \addtoversion{normal}{\mathbfit}{cmr}{bx}{it} 
  \addtoversion{bold}{\mathbfit}{cmr}{bx}{it}
  \newmathalphabet{\mathbfss} 
  \addtoversion{normal}{\mathbfss}{cmss}{bx}{n}
  \addtoversion{bold}{\mathbfss}{cmss}{bx}{n}
  \ifAMStwofonts
    \ifCUPmtlplainloaded \else
and
your
      \UseAMStwoboldmath
      \makeatletter
      \new@mathgroup\upmath@group
      \define@mathgroup\mv@normal\upmath@group{eur}{m}{n}
      \define@mathgroup\mv@bold\upmath@group{eur}{b}{n}
      \edef\UPM{\hexnumber\upmath@group}
      \new@mathgroup\amsa@group
      \define@mathgroup\mv@normal\amsa@group{msa}{m}{n}
      \define@mathgroup\mv@bold\amsa@group{msa}{m}{n}
      \edef\AMSa{\hexnumber\amsa@group}  
      \makeatother
      \mathchardef\upi="0\UPM19
      \mathchardef\umu="0\UPM16
      \mathchardef\upartial="0\UPM40
      \mathchardef\leqslant="3\AMSa36
      \mathchardef\geqslant="3\AMSa3E
    \fi
  \fi
\fi 
   
\ifnfsstwo
  \DeclareMathAlphabet{\mathbfit}{OT1}{cmr}{bx}{it}
  \SetMathAlphabet\mathbfit{bold}{OT1}{cmr}{bx}{it}
  \DeclareMathAlphabet{\mathbfss}{OT1}{cmss}{bx}{n}
  \SetMathAlphabet\mathbfss{bold}{OT1}{cmss}{bx}{n}
  \ifAMStwofonts
    \ifCUPmtlplainloaded \else
      \DeclareSymbolFont{UPM}{U}{eur}{m}{n}
      \SetSymbolFont{UPM}{bold}{U}{eur}{b}{n}
      \DeclareSymbolFont{AMSa}{U}{msa}{m}{n}
      \DeclareMathSymbol{\upi}{0}{UPM}{"19}
      \DeclareMathSymbol{\umu}{0}{UPM}{"16}
      \DeclareMathSymbol{\upartial}{0}{UPM}{"40}
      \DeclareMathSymbol{\leqslant}{3}{AMSa}{"36}
      \DeclareMathSymbol{\geqslant}{3}{AMSa}{"3E}
    \fi
  \fi  
\fi 
      
\ifCUPmtlplainloaded \else
  \ifAMStwofonts \else 
    \def\upi{\pi}
    \def\umu{\mu}
    \def\upartial{\partial}
  \fi
\fi

\title{Power density spectrum of NGC 5548 and the nature of its 
      variability}
\author[B. Czerny, A. Schwarzenberg-Czerny, Z. Loska]
       {B. Czerny$^1$, A. Schwarzenberg-Czerny$^{1,2}$, and Z. Loska$^1$\\
        $^1$N. Copernicus Astronomical Center, Bartycka 18, 00-716 Warsaw, 
        Poland\\
	$^2$Astronomical Observatory of Adam Mickiewicz University, 
   ul. Sloneczna 36, 60-286 Poznan, Poland}

\begin{document}

\maketitle

\begin{abstract}
We derive power density spectra in the optical and X-ray band in the
timescale range from several years down to $\sim $ a day. We suggest that
the optical power density spectrum consists of two separate components:
long timescale variations and short timescale variations, with the
dividing timescale around 100 days. The shape of the short timescale
component is similar to X-ray power density spectrum which is consistent
with the interpretation of short timescale optical variations being caused
by X-ray reprocessing.
We show that the observed optical long timescale variability  is
consistent with thermal pulsations of the accretion disc.

\end{abstract}


\section{Introduction}

Systematic multiwavelength study of radio quiet Active Galactic Nuclei (AGN) 
led to the emergence of the general picture of such a nucleus as a massive 
black hole surrounded by a relatively cool accretion disc of an outer radius 
below $\sim 1$ pc and a hot optically thin plasma
(see Mushotzky, Done \& Pounds 1993). Comptonization of the soft photons by 
the plasma leads to emergence
of strong X-ray emission from the nucleus. 

This 
compact unresolved region is surrounded by a dusty torus (see Antonucci 1993),
marginally resolved in the IR at the distance scales below 100 pc 
in the case of the nearest AGN (see e.g. Greenhill et al. 1996 for water maser 
emission from NGC 1068) although not in the case of NGC 5548.  

The models of the compact innermost region have to reproduce the
results of the 
detailed studies 
of the spectra of Seyfert~1 galaxies. X-ray observations were particularly 
successful in demonstrating the coexistence of the hot and cold gas close 
to the 
black hole through the detection of the reflection component (Pounds et al. 
1990) and 
subsequent identification of the observed 6.4 keV iron line profile in a 
number 
of sources with the reflection from relativistic accretion disc 
(Mushotzky et al. 1995 for IC 4329A and NGC 5548, Tanaka et al. (1995) for 
MCG-6-30-15, Yaqoob et al. 1995 for NGC 4151,  Yaqoob et al. 1996 for 
NGC 7314, Nandra et al. 1997a for NGC 3516 and Weaver et al. 1997 for MCG -5-23-16; 
for a general discussion see 
Nandra et al. 1997b).  

The shape of the high frequency tail of the X-ray spectra shows that 
the velocity distribution of electrons in the hot plasma is predominantly 
thermal (e.g. Gondek et al. 1996 and Zdziarski et al. 1997  for averaged 
Seyfert 1 spectra, and Zdziarski, Johnson \& Magdziarz 1996 for NGC 4151).

However, the dynamics of this multi-phase medium feeding the black hole is 
still not understood although a number of attempts to construct a complete -
usually stationary - model were undertaken (e.g. Narayan, McClintock \&
Yi 1996, Witt, Czerny \&
\. Zycki 1997). The best way to learn which 
processes might dominate the flow is to study the variability of the nucleus 
since the variations can be directly related to dynamics whilst the spectra 
may 
be insensitive to dynamical parameters.  

Extensive studies of variability were done so far mostly for galactic X-ray 
sources, including black holes. In the case of AGN, most of the studies
were constrained to the observations of the spectral changes based on a few
spectra and to the studies of the time delays between different spectral 
bands. Nevertheless those studies of a few monitored AGN brought important 
conclusions: the well correlated
variability in the optical, UV, EUV and X-ray band and the lack of
measurable delay between those frequency bands suggested that the optical/UV
variability is caused by the reprocessing of the variable X-ray flux
(e.g. Collin-Souffrin 1991,  Clavel et al. 1992).

However, detailed studies of that hypothesis were difficult. Power spectra
were constructed only for the X-ray data and early results based on long
EXOSAT lightcurves showed that the PDS was a featureless power law without
any characteristic timescale (McHardy \& Czerny 1987 for NGC 5506, 
Lawrence et al. 1987 for NGC 4051).

The situation is changing during the last years. Due to the undertaken monitoring 
of several AGN more and more data in different wavebands are available. It
is therefore important to study the physical constraints to the models 
imposed by these data.

It is also important to compare the results of variability studies for AGN
to what we know about the variability of galactic black holes.
The question is the possibility of 
scaling of some, or most, properties of the flow with the mass of the black 
hole which may extend over orders of magnitudes, from $10^{10}$ of solar masses 
down to ten. If such a scaling exists for some 
properties, the difference in the timescales of the AGN and galactic black 
holes help to study them in a complementary ways. Long evolution is easier to 
study for galactic sources, as the timescales of years for galactic sources 
correspond to millions of years for AGN. On the other hand the timescales 
shorter than the travel time through the horizon of a black hole, of order of 
tens or hundreds of seconds for AGN and of orders of tens of  microseconds for a 
galactic black holes are easier to reach for AGN.  

In this paper we derive the power density spectrum of NGC 5548 in the optical
band based on the data presented by Sergeev et al. (1997) including 
the data collected by AGN Watch team. We extend the PDS towards high
frequencies using the nonlinear prediction method. We estimate the broad
timescale PDS in X-ray band and we compare it to the optical one. We 
compare these distributions to the present knowledge of other AGN and of
galactic X-ray sources. Finally, we discuss the physical mechanisms of
the observed optical variations.

\section{The observational data for NGC 5548}

The nearby Seyfert galaxy NGC 5548 (z=0.0174) is the most extensively observed 
AGN. The optical/UV variability of this object was followed systematically by 
International AGN Watch (Peterson et al. 1991; Peterson et al. 1992; Peterson 
et al. 1994; Korista et al. 1995) during the last five years.  The X-ray 
variability was recently studied  by Nandra \& Pounds (1994) and 
Clavel et al.  
(1992) in the hard X-rays, by Done et al. (1995) in soft X-ray band, and new 
results are expected from XTE. The EUVE variability was detected by 
Marshall, Fruscione \& Carone (1995) (see also Marshall et al. 1997). 
These data allowed to formulate first 
phenomenological  models of the time behaviour of the flow in this object 
(e.g. Rokaki, Collin-Souffrin \& Magnan 1993; Loska \& Czerny 1997: 
Magdziarz et al. 1997).  

\subsection{Optical data}

In this paper we use the data obtained by AGN Watch team 
(Peterson et al. 1994) together with  
data accumulated over the past ten years prior to the AGN Watch activity 
and made recently publicly available (Sergeev et al. 1997).

The entire optical data set covers the period from 1977 to 1993. Although the 
calibration of the data may not be very accurate (see Sergeev et al. 1997) the 
amplitude of the variability is large (close to factor 3) and small 
calibration errors of order of 3 percent  cannot influence the general 
picture.  

Basic properties of this lightcurve become apparent by simple visual 
inspection. The variations cover all timescales but the amplitude of variations 
is related to the timescale. In the shortest timescales, day-to-day changes do 
not exceed several percent. Within one year, large variations of order 
of 40 percent peak-to-peak occurred. However, the lightcurve appears flat over the 
whole interval of 17 years, indicating possible deficit of fluctuations with 
longest time scales exceeding several years.  

\subsection{X-ray data} 

The X-ray power density spectrum was already obtained by 
Papadakis \& Lawrence (1993).
It was derived only on the basis of relatively short 
observations made by EXOSAT satellite 
which covered only
timescales below 0.3 d (more precisely, between $ 3 \times 10^4$ and 
$5 \times 10^2$ s), beyond the range covered by the
optical data. The PDS was best described as a power law with additional QPO
type feature at $\sim 10^3$ s. The slope of the aperiodic variability was not
determined accurately even for the longest data set from July 1984 
($\alpha = 1.99 \pm 0.9$). The 
variations in the 2-10 keV flux in this EXOSAT data of the length of 1.006 d 
reach the value $  0.42 \times 10^{-11}$ erg/s/cm$^2$ 
(Kaastra \& Barr 1989).

We will try to estimate the X-ray PDS at longer timescales using the
following information about the characteristic variability amplitudes. 

The variations observed by Ginga during the
42 days of monitoring reached the amplitude $ 1.58 \times 10^{-11}$erg/s/cm$^2$
(Clavel et al. 1992). The strong brightening of the source in the
optical band in 1984 was accompanied by only a moderate brightening in
X-rays showing that the amplitude of variations in timescales of 
$\sim 6$ years increases only up to $1.73 \times 10^{-11}$erg/s/cm$^2$. 

\section{Results}

\subsection{The optical power density spectrum for NGC 5548}

\begin{figure}
\epsfxsize = 80 mm \epsfbox[50 180 560 660]{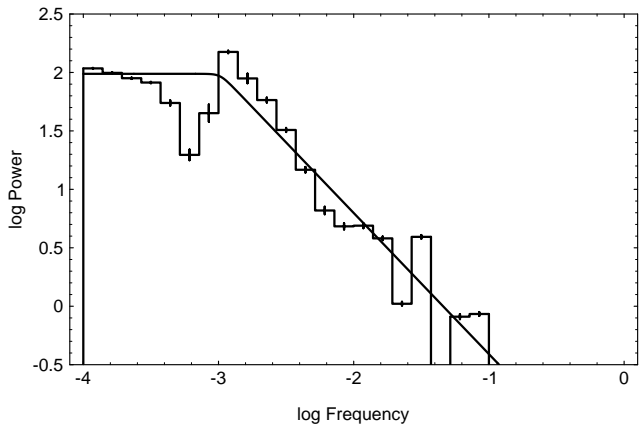}
\caption{The smoothed and binned power density spectrum  
of NGC 5548 in the optical band plotted in the decimal
log-log scales. On the horizontal axis frequency is in units of
cycles per day. On the vertical axis power density is in units of
$10^{-30}(erg/cm^2s \AA)^2 d$. Continuous line is a broken power law fit
to the power spectrum, with the slope 1.17.
}
\end{figure}

We study the optical lightcurve of NGC 5548 using the Fourier analysis method. 
Although a number of other techniques were applied to study variability, 
determination of the power density spectrum still remains the basic 
approach even in the study of aperiodic variability (e.g. van der Klis 1995).
  
In the case of AGN only X-ray (and not the optical) variability was 
studied in such way, with 
exception of a BL Lac object OJ 287 (Silanp\" a\" a et al. 1996), 
since the time extension and the coverage by 
optical data is usually not good enough to perform that kind of analysis.  

We calculate the power density spectrum (PDS) using an algorithm of Deeming (1975) and applying
sufficient sampling interval of 0.00001 $d^{-1}$. In order to 
reduce the noise and to enhance the signal we smooth
the power spectrum with a median filter and then bin it on the log scale.  
Following Korista et al. (1995) we adopt such normalization, that PDS
is computed in units of flux squared times day, 
i.e. $10^{-30}(erg/cm^2s\AA)^2d$. At redshift $z=0.0174$ the adopted unit
of PDS corresponds to power in B band of $30 \times 10^{18} L_{B\odot}^2d$ 
(c.f. Korista et al., 1995). No correction for power leaks to
high frequency induced by window function
was applied. Simulations demonstrate that leaks are critical for
spectral indices approaching -2 (Green, McHardy \& Lehto 1993). Major source of uncertainty
in our power spectrum are low frequency fluctuations for which statistics 
in our data is
poor and which contribute large fraction of total power in 
the power spectrum.


\begin{figure}
\epsfxsize = 80 mm \epsfbox[0 500  400 800]{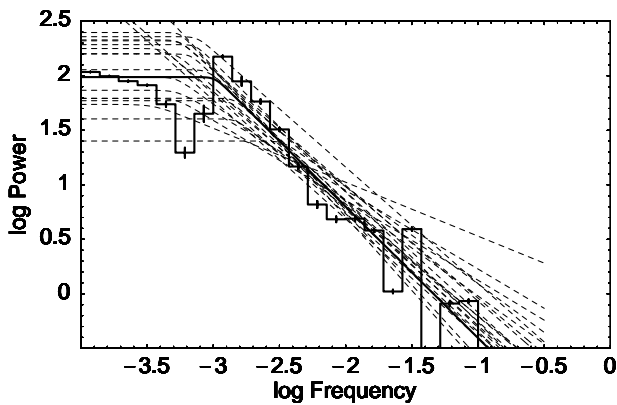}
\caption{The smoothed and binned power density spectrum  
of NGC 5548 given as in Fig. 1, together with a direct fit (continuous line).
Dashed lines show  examples of numerical simulations probing the influence
of the window function onto the fit. 
}
\end{figure}

The resulting PDS is shown in Fig. 1. It is plotted in 
log-log axis usually adopted in X-ray variability analysis. Timescales 
shorter than $\sim 30$ days are dominated by white noise and better time resolution 
is required to study that range. We subtracted this constant component
from the whole spectrum before plotting. 
The overall spectrum 
is dominated by clearly aperiodic signal. It can be described as 
a power law with index $\sim 
-1.17\pm0.04$ flattening off at time scales longer than about 3 years.  
We show the fit of such a power density shape 
to the observed one (Fig.1, continuous line). 

The above error estimates base on the internal scattering of the residuals
from the fit. The influence of the window function itself on the shape of the
spectrum was analysed by Papadakis \& Lawrence
(1995). They conclude that the power leak to high frequencies is negligible
for inclinations flatter than -2. 
Our own extensive simulations (Schwarzenberg-Czerny et al, in
preparation) show that the 
actual external error of the slope may be as large as 
0.2. However, the simulations confirm the existence of the break - its 
frequency is reproduced with accuracy $log \nu_o = 3.2 \pm 0.3$. 
Exemplary results of such simulations are shown in Fig. 2.

At lowest frequencies our PDS is affected by smoothing. Hence we performed independent analysis of low-frequency
variations. For this purpose we splitted data into two parts, namely AGN Watch data and pre-Watch data and then computed moments for each part separately.
Variances in two parts are identical, corresponding to standard deviation of
individual observation of $\sigma=1.63$. Corresponding averages, namely
8.61 and 10.05 differ by less than a standard deviation. This may mean that 
PDS at lowest frequencies does not exceed average PDS value, in agreement with
Fig.1. Note that because
of correlation of individual observations, we expect difference of averages
considerably in excess of white noise expectation $\sigma/\sqrt{(n-1)}$. 
This argument, as model independent, is stronger than any results from the 
simulations. Our simulations, however, support this conclusion 
(Schwarzenberg-Czerny et al, in preparation). 

Therefore, the flattening of the optical PDS at low frequencies is
a real property of the data. However, we cannot tell from this
analysis whether actually 
the change of slope is rapid, or gradual. The results of the nonlinear
prediction method seem to support the second possibility (see Section 3.2).

Formally, our PDS indicates the presence of a periodicity, or 
quasi-periodicity, at a knee of the basic shape (i.e. $\sim 2.5$ years), and 
its 
harmonics of 1.2 years. These features are still better seen before 
smoothing and binning of the PDS.
However, the significance of that feature is 
doubtful for several reasons. On the one hand, the whole data cover only 7 
periods for such a timescale. The power at this time scale comes from
the two minima observed in  1990 and 1992 as the PDS computed 
for data excluding this minima does not display this feature.
Any process generating large fluctuations would produce such PDS
features if by chance exactly two of them occurred during the
observed interval. Let us assume for the moment that the probability of observing minima
in the lightcurve of NGC5548 is uniform in time.
Then, the probability of finding two of them within 2.5 years during the
17 years of observations is as high as
$2.5/17\sim 0.15$.

\subsection{The results of the nonlinear prediction method for NGC 5548}


\begin{figure}
\epsfxsize = 80 mm \epsfbox[50 180 530 660]{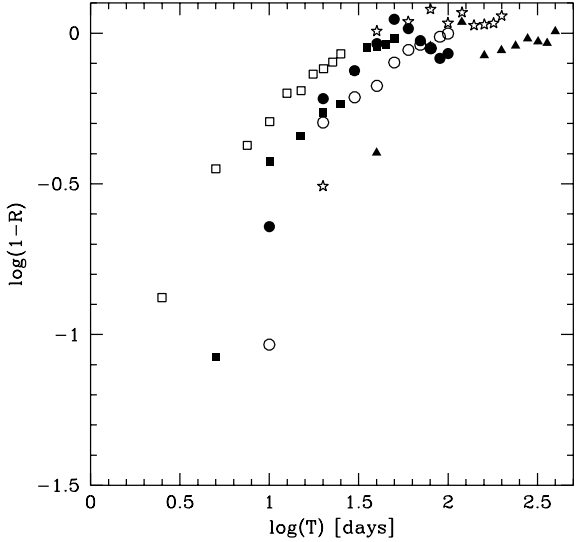}
\caption{The quality of prediction as a function of the prediction time-step
for different timestep of the interpolated data: open squares (2.5 d, AGN
Watch data), filled squares (5 d, AGN Watch data), open circles (10 d, AGN Watch), filled circles (10 d, whole data), stars (20 d, whole data), filled
triangles (40 d, whole data) 
}
\end{figure}

\begin{figure}
\epsfxsize = 80 mm \epsfbox[50 180 530 660]{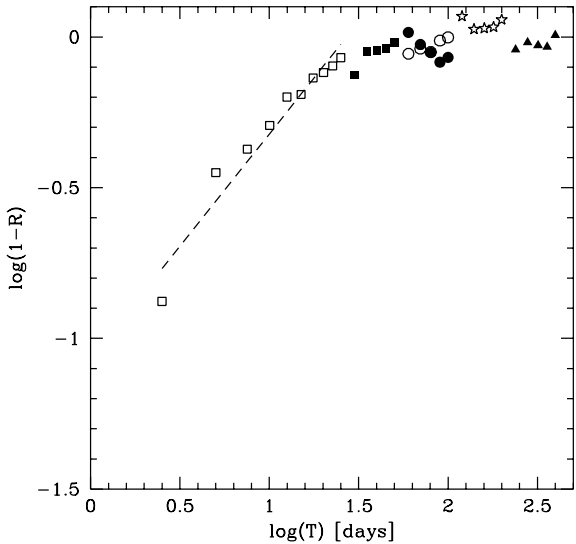}
\caption{The quality of prediction as a function of the prediction time-step:
the combined picture. Data for prediction time below 25 d are from 
interpolation 2.5 d set, between 20 d and 50 d from interpolation 5 d set,
from 50 d to 100 d from interpolation 10 d set (for AGN Watch data 
and, separately, from whole data), between 100 d and 200 d from interpolation
20 d set and finally between 200 and 400 d from interpolation 40 d set. 
The correlations for the longest prediction time steps are close
to zero.
The dashed line shows the linear fit to the high frequency part well
approximated by a power law. The slope of this power law equal to 0.74 
corresponds to the PDS slope 1.74. 
}
\end{figure}

The high frequency behaviour of the data is better analysed with the use
of the nonlinear prediction method which is sensitive to the timescales as
short as the typical separation between the measurements.
This method, introduced by Tsonis \& Elsner (1992) and successfully applied to 
X-ray data for AGN by Czerny \& Lehto (1997) is based on the quantitative
analysis of the decay of the correlation between any two short sequences in the
data with an increase of the time separation of those sequences. It does not
require any subtraction of the white noise.

The correlation coefficient R is usually a decreasing function of the  time
separation T. If the signal is well represented by a simple shot noise, 
its PDS is well described by a power law with an index $\alpha$ and at the
same time the
functional form of the dependence R(T) takes the shape 
\begin{equation}
1 - R(T)  \propto T^{\alpha - 1}.
\end{equation}

Therefore, in the case of non-periodic signal with basically a power law
shape of PDS the nonlinear prediction method can be used to determine
independently the slope of the PDS.

The method was developed for the equally spaced data with gaps so the
original data had to be interpolated. 

To study the shortest timescales we used the data collected by 
AGN Watch team. The mean separation of the data points is 3.4 days,
with sequencies sampled as densely as 1 day during the fifth year so
we adopted the timestep 2.5 d for interpolation. We included only
these data points which were interpolated between the original
data points separated not more than 1.6 times the adopted timestep.

The resulting correlation coefficient $R$ between the actual and
predicted values of the flux as a function of the prediction time $T$
is shown in Fig. 3 (open squares). The log(1-R) vs. log(T) plot 
is generally of a power low shape although a conclusion about 
the presence of some
curvature cannot be rejected.  The slope of this power law
is equal 0.74 which translates into the slope of the PDS equal 1.74
 between
the frequencies 0.04 and 0.4 $d^{-1}$.

The signal is clarly of stochastic 
nature and not od deterministic chaos since the linear fit to
log(1-R) vs. log(T) is much better (the correlation coefficient equal 0.95) 
than the linear fit to log(1-R) vs. T diagram (the correlation 
coefficient equal 0.83).

We repeated the analysis assuming
the interpolation timestep equal 5 and 10 d (filled squares and open circles)
which provided results for longer timescales. The slopes do not
differ more than by 0.1 from the previous case although for longer
timescales the slope is determined with larger error. Also the
statistics of the pairs for prediction becomes a problem. Therefore
we repeated the analysis for 10 d interpolation timestep using the
whole data, i.e. including the entire period of 17 years of 
observations (Fig. 3, filled circles). We also attempted to detect the
long timescale trends using 20 d and 40 d interpolation timestep but
these results are already very close to R=0 line (no correlation).

In all computations the dimension of the string $d$ 
(i.e. the length of the data sequence used
for prediction; see Czerny \& Lehto 1997)
was obtained from the criterion that the correlation for prediction
over a single timestep is the largest. In all cases $d$ was equal 1.    

As the best representation of the data we choose the results obtained
from 2.5 d interpolation step, supplemented at longer timescales
by the computations made with the use of consecutive longer interpolation
timesteps (Fig. 4). However, it is difficult to make any definite
statement about the flattening of the curve at timescales about 
30 d since
the correlations at those timescales become very weak.

To see better the comparison between the results based on nonlinear
prediction method and on classical Fourier analysis we replot both
on a single plot using the proportionality: 
\begin{equation}
PDS \propto (1-R(T))T
\end{equation}
and adjusting the normalisation to the Fourier results (Fig. 6).
The overlapping region is large enough to see that the results 
of the nonlinear prediction method extend nicely the Fourier derived
PDS to high frequencies without any sudden change of the slope. 
However, a better data would be necessary in order to see whether
a step like feature at $\sim 50$ days is present in the optical
data or rather the PDS is well represented by a single component
with a significant curvature. Further work of AGN Watch team may
resolve this problem.

\subsection{X-ray power density spectrum}

\begin{figure}
\epsfxsize = 80 mm \epsfbox[50 180 530 660]{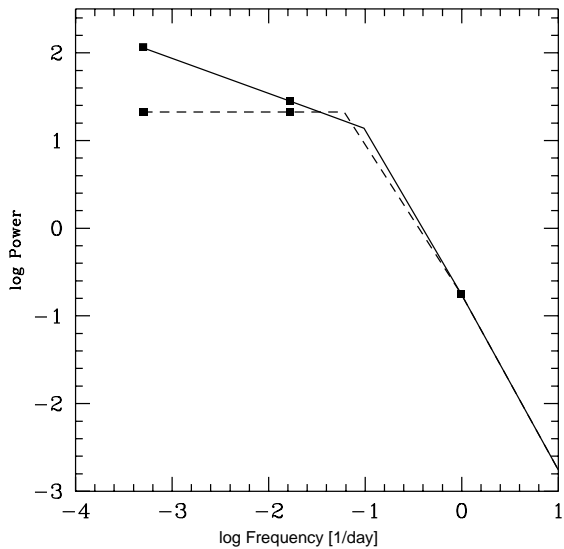}
\caption{The X-ray power density spectrum in 2 - 10 keV in units $10^{-22}
(erg/cm^2/s)^2 d$. The dashed line shows the solution with adopted 
low frequency slope equal zero and continuous line shows the solution
with the steepest
possible slope in low frequency band (see Section 3.3). Points mark the frequencies at which the amplitude
was measured. }
\end{figure}

In view of lack of a complete X-ray luminosity curve covering all 
relevant time scales we consider maximum range of variability at a given 
timescale as more robust measure of variance.  Given an estimate of the 
variance at several time scales, we proceed to estimation of power 
density function $P(\nu)$ using Parseval equation, 
  \begin{equation}
Var(\nu_o) = {\int^{\infty}_{\nu_o} P(\nu)d\nu} \label{e:111}
\end{equation}
which reduces the problem to the inversion of an integral equation.
We defer derivation of Eq. (\ref{e:111}) and its inversion to Appendix.

At the shortest timescales (below 0.68 day) we adopt the PDS slope equal
1.99 after Papadakis and Lawrence (1993). We normalize it on the basis
of equation (2) using the
variability amplitude for the 2-10 keV flux in the EXOSAT data given 
by Kaastra \& Barr (1989).

We assumed that PDS between $\nu=1 d^{-1}$ and $1/6 yr^{-1}$ may be
represented by a broken power law with a break at $\nu_o$. The parameters of 
the power law are determined by inversion of Eq. (\ref{e:111}), or more 
specifically of its particular form (see Appendix, Eq. \ref{e:112}). 

In the first solution we assumed $\alpha_2=0$. 
Our solution yields break frequency $\nu_0 = 6.0 \times 10^{-2}d^{-1}$ 
and the slope $\alpha_1 = 1.71$ (see Fig. 5, dashed line).
The position of the $\nu_o$ depends very strongly on the difference
between the amplitudes at $\nu=1/42 d^{-1}$ and $1/6 yr^{-1}$ and the last value is actually given by a single measurement so the derived value should be treated as an estimate of the order of magnitude. 
We also constructed solutions with arbitrary but nonvanishing value
of $\alpha_2$. The solutions existed only if $\alpha_2$ was smaller
than 0.41. We show this limiting case in Fig. 5 (continuous line).

We are unable to exclude negative values of $\alpha_2$,
for our method does not work well for non monotonic PDS.

The two limiting cases for the solutions of the constraints imposed
on X-ray PDS well illustrate the overall character of the X-ray
variability. Clearly, the PDS cannot be a single power law 
in the entire studied timescale range. However, we cannot tell
whether the turnoff is sharp or rather we deal with a broad band
curvature.

\section{Discussion}

\subsection{Character of the optical variability}

\begin{figure}
\epsfxsize = 80 mm \epsfbox[50 180 530 660]{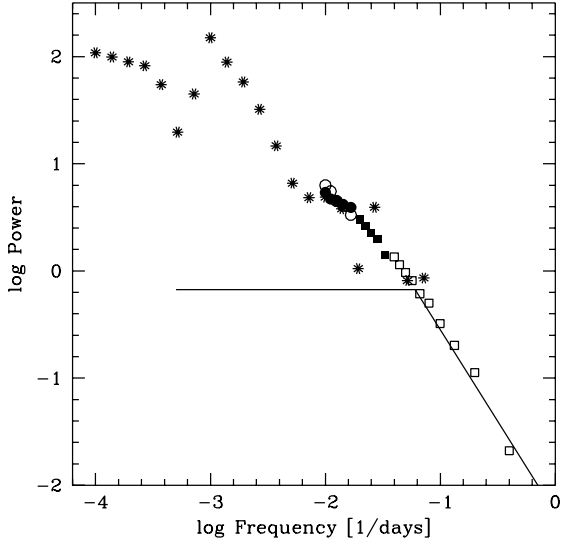}
\caption{The combined information about the power density spectrum of
NGC 5548: stars are the optical PDS from Fig. 1, points of various shape
mark the optical PDS derived from the nonlinear prediction method 
(Fig. 4) for
timescales shorter than 100 d,  continuous line mark the
X-ray PDS from Fig. 5. The normalization of the last two components were
adjusted to match the high frequency tail of the first component.
}
\end{figure}

\begin{figure}
\epsfxsize = 80 mm \epsfbox[50 180 530 660]{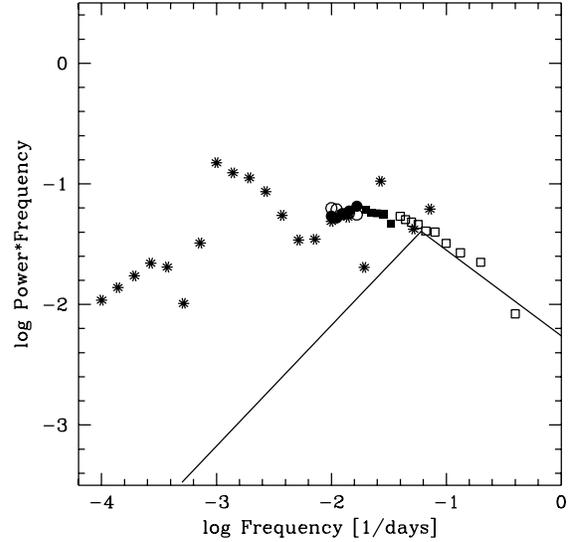}
\caption{The same as Fig. 6 but after multiplication by the 
frequency to show the power per logarithmic unit of the frequency;
the peak on this diagram shows the frequencies at which most of the 
power resides.}
\end{figure}

The optical variability is aperiodic and broad band covering all the available
timescales from years down to a day. Present data do not allow to distinguish
any QPO phenomena superimposed on the broad component  although the traces
of some substructure in the PDS seem to show up.

The overall variability is dominated by the timescales around 3 years. The
PDS would show it clearly if reploted as power per frequency decade
i.e. $log(Power \times Frequency)$ 
versus
$log(Frequency)$ diagram, in analogy to popular representation of the 
radiation 
flux as $log(Flux \times Frequency)$ vs. $log(Frequency)$. 
In such a new plot the
flat part of the PDS  change into an steeply increasing power law 
peaking at a turn off point at 3 years (See Fig. 7).

The subsequent part, between the timescales 
from $\sim 3$ years down to $\sim 100$ days looks almost flat since the
power law index 1.17 is translated into just 0.17 in the new diagram
so the PDS is really broad band. This part ends up possibly with a 
second peak at timescales of about 30 d. 

We also cannot reject the hypothesis that the PDS shows a continuous
curvature with a single broad flat part covering the timescale
range between $\sim 30 d$ and $\sim 3 y$. 

At the shortest timescales (below 30 days) the decrease of the variability amplitude is
much faster, with PDS index equal 1.74. 

The data may therefore represent a single variability mechanism operating
most efficiently between the timescales of 3 years and 30 days.

However,
a flattening of the PDS at $\sim 200$ days with subsequent turn off at
$\sim 30 $ days may rather suggest that the optical
PDS consists of two separate components.
The long timescale component has a flat PDS above 3 years and turns down
with an index somewhat steeper than the overall 1.17. At the shortest 
timescales there is an onset of another component which again is flat
above 100 days, peaks somewhere at about 30 days and then turns off rather 
steeply.

\subsection{Comparison with X-ray variability of NGC 5548}

The dominant timescale in X-rays is somewhere around 17 days,
although may be a factor of a few longer if the amplitude at 6 year 
timescale
is underestimated. This is clearly different from the
timescale of 3 years dominating the optical variability. However, it
coincides interestingly with some substructure present in the optical PDS
at the timescale of $\sim $ 100 days.

Particularly, if the optical PDS consists of two components, as suggested
in Section 4.1, we can roughly identify the X-ray PDS shape with the
short timescale component. In both cases the PDS is flat above 100 days,
the turnover is around $\sim 30 $ days (30 in Fourier approach, 30
in nonlinear prediction and 17 in X-ray data) and the slope after the
turn-off is steep (-1.71 for the timescales of days in X-rays, -1.74
in optical band).

In order to show better the comparison of the optical and X-ray PDS
we placed the X-ray PDS on the same plot as the optical one
(see Figs. 6 and 7). Since in that
case units are clearly different we adopted the normalization of the 
X-ray PDS in that case as arbitrary and we adjusted it to the 
high frequency tail of the optical PDS. The deficit of the power
in X-ray PDS with respect to the optical PDS at longer timescales
are now clearly seen.

Therefore the data seem to suggest the possibility that on the timescales
longer than $\sim 100 $ days the optical variability is unrelated to X-ray
variations while on the timescales  below $\sim 100 $ days the shape of the 
optical PDS is dominated by another component of the shape similar to 
X-ray PDS.

\subsection{Comparison with X-ray variability of AGN}

X-ray PDS for other AGN are determined more accurately. 
For us the most interesting cases are two Seyfert galaxies studied in the 
broad frequency range.
 
NGC 5506 is a bright Seyfert 2 galaxy. The long history of the X-ray 
measurements
of that source together with a 3 day continuous observation by EXOSAT 
allowed to determine its PDS in the entire range from
$\sim 10$ years down to 400 s (McHardy 1989). The shape of the PDS  is
basically that of 
a power law with index $\sim 1.5$, clearly flattening
above the timescales of order of $\sim 20$ days. Its high frequency part
between $10^{-5} - 10^{-3}$Hz was actually better represented as a broken 
power law changing index from 1.0 to 1.9 towards high frequencies.  

The broad band X-ray PDS spectrum of NGC 5548 derived by us is therefore
quite similar to NGC 5506. Even the position of the $\nu_{flat}$ is 
similar which may not be surprising since the luminosities of the two
objects are comparable.

NGC 4151 is a nearby moderately bright Seyfert 1.5 galaxy frequently observed
in X-rays. The broad band PDS for that galaxy also shows clearly the flat
part at timescales $\sim 100 $ days and a power law with index $\sim 2.1$ 
between $10^3$ and  $5 \times 10^4$ s (Abraham \& McHardy 1989; see
also Papadakis \& McHardy 1995). It is, 
however, impossible to tell on the basis of that data 
whether the flattening progresses continuously towards low frequencies or a 
well defined knee is present somewhere at the timescale between a day and 
several days.
 
Few other AGN were also studied
basing on the long EXOSAT lightcurves which allowed to determine accurately
the slope of the PDS for timescales between fraction of an hour and a day.
Mean index of that slope determined from a sample of a few objects was 
$1.7 \pm 0.5$ (Green et al. 1993) or $1.55 \pm 0.09$ (Papadakis \& Lawrence
1993). Studies based
on Ginga data (e.g. McHardy 1989) as well as complementary studies of 
the old EXOSAT 
lightcurves made using the non-linear 
prediction method (Czerny \& Lehto 1997) showed that
power law continues down to the timescales of 200 s.  Recent study of the ASCA
data for MCG 6-15-30 with the use of excess pair fraction method (Yaqoob 
et al. 1997)
indicate that the power law with an index $\sim 1.5$ extends actually down
to 20 s timescale.  However, no significant change in the slope of the power
law was noticed.


Certain degree of success in modelling X-ray lightcurves of AGN
was achieved using Linear State Space Models employing
first order autoregressive process AR[1] for modelling of dynamics
(K\"{o}nig \& Timmer, 1997). This model accommodates flattening of 
the spectrum both at low and high frequencies.


\subsection{Possible mechanisms of optical variability}

The optical variability is dominated by the broad band of timescales
starting from 3 years and extending down to shorter timescales with only
slowly weakening power.

This timescale is too short for typical viscous evolution which requires 
thousands of years (see e.g. Siemiginowska, Czerny \& Kostyunin 1996). 
Therefore we look for explanation of the observed behaviour amongst the
dynamical or thermal timescales.

\subsubsection{Dynamical instabilities}

The basic dynamical timescale is given by the Keplerian orbital motion.
The mass of the black hole in NGC 5548 is most probably within the range
$6 \times 10^7$ and $1.4 \times 10^8 M_{\odot}$ (see e.g. Loska \& Czerny 
1997, Kuraszkiewicz, Loska \& Czerny 1997 and the references therein). 
Therefore, we can adopt the
value  $10^8 M_{\odot}$ as representative.

The dynamical timescale is equal 3 years when the distance $x$ (expressed in
units of the Schwarzschild radius) is equal
\begin{equation}
x= 1600 M_8^{2/3}.
\end{equation} 
This distance translates into a distance of 16 light days for $10^8 M_{\odot}$
black hole. 

If the disc does not extend beyond that radius and it is clumpy or dynamically
pulsating below it then the observed range of timescales is reproduced.

There are no easy arguments why the disc should be truncated at
that radius. However, low ionization  broad emission lines (LIL), 
if formed at the
disc surface, (e.g. Collin-Souffrin  1987) show the typical delay of 18 
days (Peterson et al. 1994), i.e. about the required light travel time.
Also the characteristic timescale for the X-ray variability and the short
timescale optical component peaks somewhere around this value which may
correspond to the light travel time across the hot plasma region but
those two timescales can be also related in a natural way without
any reference to the dynamical timescale in the disc (see
Section 4.6).

\subsubsection{Thermal instabilities}

The thermal timescale at any given radius depends on the assumptions about the
properties of an accretion disc. If we assume that the disc structure is well
represented by the classical model of Shakura \& Sunyaev (1973) then the
thermal timescale 
is longer than the dynamical
timescale by a factor $\alpha ^{-1}$, where $\alpha$ is the viscosity
parameter. The location of the radius where the
thermal timescale is equal to 3 years depends on the adopted value of the
viscosity parameter.

The only value suggested for that particular object is 0.03 derived on the
basis of the overall optical/UV/X-ray spectra (Kuraszkiewicz et al. 1997) 
within the frame of the accreting corona model. Adopting that value for
a scaling we obtain the radius
 
\begin{equation}
x= 160 \alpha_{0.03}^{2/3} M_8^{2/3}
\end{equation} 
which is at the distance of 1.6 light day from the central black hole of
the mass $10^8 M_{\odot}$.

The inner parts of AGN accretion discs, within a frame of the $\alpha$ times 
the total pressure viscosity model, are thermally unstable.
We can compare the radius given by equation 4 with the radius where there
is the onset of the instability.

Simple estimate of the transition radius from the radiation pressure to gas
pressure domination gives (Shakura \& Sunyaev 1973)
\begin{equation}
x_{ab}= 46 \alpha_{0.03}^{2/21} M_8^{-2/3} \dot M_{0.025}^{16/21}
\end{equation}
where we adopted the mean accretion rate 0.025 $\dot M_{\odot}$/yr (after
Kuraszkiewicz et al. 1997) as a
scaling factor.

\begin{figure}
\epsfxsize = 80 mm \epsfbox[50 180 530 660]{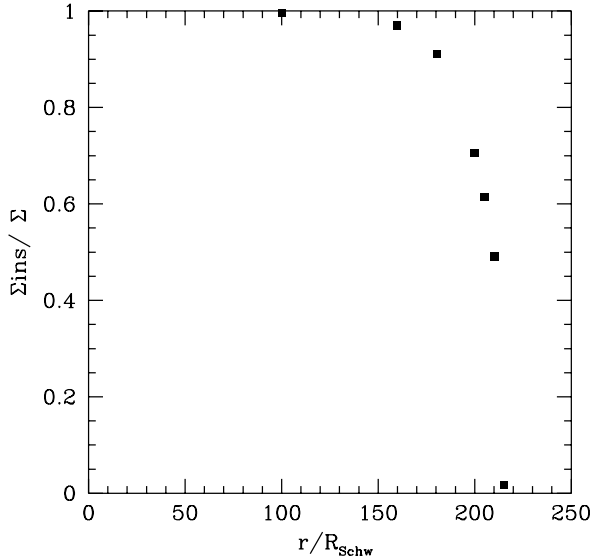}
\caption{The fraction of the disc which is thermally unstable plotted as
a function of radius. Model parameters: $M=10^8 M_{\odot}$,$\dot M = 0.025
M_{\odot}$/yr and $\alpha = 0.033$. 
}
\end{figure}

This value only weakly depends on the viscosity and is by a factor $\sim 3$
smaller than expected observationally. However, more accurate computations
are encouraging. 

Our accretion disc model was computed taking into account the vertical 
structure of the disc (R\' o\. za\' nska et al. 1998). 
Free-free opacity was replaced by the appropriate
full opacities (Kramers means) in the radiation transfer equation.  

The disc is thermally (and viscously) unstable when the ratio $\beta$ of the
gas pressure to the total pressure is smaller than 2/5 (Shakura \& Sunyaev 
1976). Since we now compute the vertical structure (so the gas to the 
total pressure ratio is also a function of the distance from the equatorial 
plane) we estimate the tendency of the disc to develop the thermal 
instability by computing, at any radius,  the ratio of the disc mass 
where $\beta$ is smaller than 2/5 to the total mass. 

In Fig. 8 we show the plot of the fraction of the disc mass which is
thermally unstable at a given radius. It is calculated as a ratio of
the surface density of the gas with $\beta$ smaller than 2/5 to the total
surface density.

We see that the instability starts to be present at $\sim 200 R_{Schw}$
and at $150 R_{Schw}$ the entire disc become thermally unstable so
the thermal instability of an $\alpha$ disc is a viable explanation of
the shape of the long timescale optical PDS.

\subsection{Possible mechanisms of X-ray variability}

The interesting conclusions about the nature of X-ray variability were
derived so far only for galactic sources. Recent XTE  
observations of Cyg X-1(Cui et  al.
1997) applied to the concept of coherence function  (Vaughan \& Novak 1997)
suggested that most of the time the hot medium itself is
not strongly variable. The observed variability may be caused instead by 
scattering  of the soft photons emitted close to the black hole by an 
 extended medium (Kazanas, Hua \& Titarchuk 1997). In this
picture the timescale of the flattening of the PDS measures directly 
the size of the comptonizing region.

In the case of NGC 5548 it would mean that the hot medium extends up
to $\sim 30$ light days.

One of the possibilities is that the Compton heated corona
forms above an accretion disc (Begelman, McKee \& Shields 1983). The outer
radius of such a corona is located at $\sim 10000 R_{Schw}$ (i.e.
100 light days). The corona, however, scatters photons efficiently 
only up to $\sim $ 10\% of that distance (e.g. Ostriker, McKee \& Klein
1991) which
clearly makes an interesting explanation of the presence of the
hot medium within the distance required by the shape of the X-ray PDS
although it would be necessary to invoke some extra heating in 
excess of the Inverse Compton temperature. 

The fact that the representative
emission radius of the $H_{\beta}$ is comparable to the disc fraction
covered by a corona may not be accidental as the presence of the
corona scattering the photons emitted closer to the black hole and
redirecting a part of them towards the disc surface enhances the 
emission of lines (see e.g. Kurpiewski, Kuraszkiewicz \& Czerny 1997).

\subsection{Comparison with galactic X-ray sources}

Rapid aperiodic variability occurs in all types of X-ray binaries (for a 
review see van der Klis 1995). Since the observed X-ray flux from these objects
is typically a few orders of magnitude higher than the observed X-ray flux
from AGN the shape of the PDS for galactic sources could have been studied 
in much more detail than for AGN. Therefore, in order to make an adequate
comparison we have to concentrate only on the basic charactersitics of
galactic sources.

The broad band PDS component 
consisting of a flat part below the frequency
$\nu_{flat}$ and a power law with index higher than 1.0 above it is 
characteristic for several different types of objects in different states.

Black holes clearly display that kind of variability in their low 
(hard) states, with
Cyg X-1 being the best example. The value of $\nu_{flat}$
is somewhere around 0.1 Hz but it may vary by a factor of a few between 
separate
observations (Belloni \& Hasinger 1990).   The power law part index is 
$\sim 1.1$ between  $\nu_{flat}$ and $\sim 3 $ Hz, and it $\sim 2.0$ at
high frequencies. The value of  $\nu_{flat}$ is $10^7$ times higher 
than the $\nu_{flat}$ of the optical PDS for NGC 5548, as it would be 
expected from a simple scaling with the mass of the central black hole.
This fact, together with the slope of the PDS suggests that the X-ray PDS
for Cyg X-1 in low state corresponds to the optical (and not X-ray!)
PDS for NGC 5548. Similar effect can be present in the case of NGC 4151:
X-ray PDS break at 14 d (Lawrence \& McHardy 1995) is an order of magnitude
too law to provide a perfect analogy with Cyg X-1. Unfortunately, optical
PDS is not available for this source.

What is more, the frequency of the change in the PDS slope
of Cyg X-1 is again by a factor $10^7$ higher than $\nu_{flat}$ of the
X-ray PDS of NGC 5548 thus indicating again a kind of substructure. Such
a view is confirmed by recent high quality XTE observations of that source
during the transition from the hard state to the soft state (Cui et al. 1997).
In this case the power spectrum is clearly two component, with one 
component having   $\nu_{flat}$  at 3 Hz (their Figs. 2 and 3) followed
by a steep power law and second component, an underlying power law, 
relatively flat
and without any clear flattening within the studied range of timescales 
(above 0.03 Hz).

If the explanation of the shape of the optical and X-ray power density spectrum
of NGC 5548 is correct (see Section 4.4 and 4.5) we would expect the simple
scaling with mass of the central black hole in the case of $\nu_{flat}$ from
X-ray PDS related to the size of the hot medium. The timescales related to
accretion disc also scales roughly with the mass of the central black hole but
they are at the same time considerably influenced by the luminosity to the Eddington luminosity
ratio.

The question, however, remains why the PDS of galacic Black Hole are similar
to optical instead of an X-ray PDS of NGC 5548. In the soft X-ray
range the answer may lie in the contamination of the spectrum by the
direct contribution of the disc to the X-ray emission due to much higher
temperatures of GBH discs due to the scaling as $M^{-1/4}$ with the mass
of the black hole. However, in the case of hard X-rays the explanation should
be different. It may be related to the fact that viscous timescales are 
relatively much
shorter in galactic objects than in AGN.  Therefore, in GBH both timescales
can mix since the viscous timescale is only
tens or hundreds times longer than thermal timescales. In AGN the viscous 
timescales are up to hundreds of thousands times longer than thermal 
timescales. From that point of view, the observational determination of  $\nu_{flat}$  in 
all states is crucial.

The same basic component also dominates the PDS of the island sources - low 
luminosity accreting neutron stars with negligible magnetic field (see
van der Klis 1995). The value
of $\nu_{flat}$ is in that case $\sim 0.3$ Hz, again as it would be expected from
the scaling of the optical PDS of NGC 5548. It indicates that this kind
of variability is not necessarily characteristic only for acreting black holes.

Finally, similar component is also typical for disk-fed X-ray pulsars like
Cen X-1, with the value of $\nu_{flat}$ again $\sim 0.3$ Hz.   

However, in their 
high (soft) states
black holes are characterized by a single power law PDS shape over several
decades, with an
index quite close to 1, as in the case o the optical PDS of NGC 5548 at intermediate
frequencies. If any flattening is present it has to take
place at frequencies lower than 0.001 Hz, at least in some sources. It may be too
long for thermal timescales within the thermally unstable part of the disc.

The qualitative changes are characteristic not only for black holes.
Also accreting neutron stars change their variability (and spectral) appearance
with an increase of the accretion rate. Island sources change into banana 
sources and the PDS of upper banana sources - a single broad band 
power law with index only slightly larger than 1 - strongly resamble the soft
state of Cyg X-1 (van der Klis 1995).

At this stage our comparison of variability of NGC 5548 with galactic sources
indicates both some similarities as well as differences and cannot be conclusive.

\section{Conclusions} 

Presented results support the view that there are two clearly distinct
physical mechanisms responsible for variability of NGC 5548.

In the short timescales (below $\sim 100$ days) the variability is caused by 
the Comptonization of the soft photons (emitted by the innermost part of
accretion disc, i.e. very close to the black hole) by an extended hot medium,
perhaps a corona. This scattering results both in variable X-ray flux and in
variable optical flux due to absorption of a fraction of the X-ray flux by
the accretion disc. The size of the hot gas region required by the shape of 
the power density spectrum is well estimated by the theory of the
Compton heated corona.

In the long timescales (above $\sim 100$ days) the optical variability is 
unrelated to X-rays. It is caused by thermal pulsations of the accretion disc.
The size of the region of the disc, required by the shape of the power density
spectrum, is consistent with the region dominated by the radiation pressure 
and therefore thermally unstable within the frame of a standard disc model
of Shakura \& Sunyeav (1973).

The scaling properties of the variability with mass are difficult to confirm
on a basis of a simple comparison of NGC 5548 with the galactic sources. The 
difficulties lie in the relative closeness of the viscous and thermal timescale
in galactic sources (factor 0.01 instead of 0.00001 in AGN) and the 
contamination of X-rays in galactic sources by a direct emission of an 
accretion disc due to the systematic difference in the disc temperatures 
between AGN and GBH by two orders of magnitude. More research should be done 
on even approximate extension of the X-ray PDS
of other AGN onto the timescales of years and on studies of the
shape of PDS
of dfferent galactic sources   on the timescales longer than 100 s in order to
confirm or reject the presented physical interpretations of the variability.

\bigskip

\section*{Acknowledgements}

We thank Andrzej Zdziarski for suggestion to introduce the Power*Frequency 
plot (Fig. 7) and to the referee, Ian McHardy, for helpful remarks which
allowed to improve the manuscript. This work was supported in part 
by grants 2P03D~004~10 (BCz \& ZL) and 2P03D~003~13 (ASC) of the Polish State Committee for Scientific Research (KBN).

\appendix

\section*{Appendix}

Link between variance and power density spectrum (PDS) may be 
established using Parseval equation:
  \begin{equation}
\int^{\infty}_{-\infty} |f(t)|^2dt = 
    \int^{\infty}_{-\infty} |{\cal F}(\nu)|^2d\nu \label{a:1}
\end{equation}
Let us consider model signal $f(t)$ binned in time intervals $\delta t$ 
differing from null in the observed time interval $T$ and 
vanishing elsewhere.  Signal $f(t)$ constitutes an example of a band 
limited signal. Its Fourier transform decreases outside frequency band 
$<1/T,1/\delta t>$. Taking this into account and assuming reality 
of $f$, so that $|{\cal F}(-\nu)|^2=|{\cal F}(\nu)|^2$ one obtains 
approximation of Eq. (\ref{a:1}) in the following approximate form 
  \begin{equation}
\int^{T/2}_{-T/2} |f(t)|^2dt \approx 
   2{\int^{1/\delta t}_{1/T} |{\cal F}(\nu)|^2d\nu} \label{a:2}
\end{equation}
Without loss of generality we may assume here that mean value was 
subtracted from data so that $<f>=0$. Then LHS of Eq. (\ref{a:2}) 
reduces to variance estimate $<Var(\nu)>$. Modulus of Fourier transform 
in RHS $|{\cal F}(\nu)|^2$ may be expressed by PDS $P(\nu)$, but 
normalization requires care.  We use Deeming (1975) Eq. (61) to ensure 
proper normalization.  We take into account that, for considered $\delta 
t$ and shape of PDS, integration up to $1/\delta t$ is equivalent to 
integration to infinity.  Finally we obtain Eq.  (\ref{e:111}): 
  \begin{equation}
T<Var(\nu)> = 2T{\int^{\infty}_{\nu} P(\nu')d\nu'} \label{e:1111}
\end{equation}
In this paper we consider PDS consisting of two power laws,
joined at $\nu_o$:
  \begin{equation}
P(\nu) = \left\{ \begin{array}{ll}
P_o\;(\nu/\nu_o)^{-\alpha_1}& \nu > \nu_o\\
P_o\;(\nu/\nu_o)^{-\alpha_2} & \nu < \nu_o
\end{array}\right.
\end{equation}
where $P_o$, $\nu_o$,  $\alpha_1$ and $\alpha_2$ are constants.
Then integration in Eq.  (\ref{e:1111}) may be 
carried analytically yielding: 
  \begin{equation}
<Var(\nu)> = \left\{ \begin{array}{cc}
\frac{2P_o\nu_o}{1-\alpha_1}\left(\frac{\nu}{\nu_o}\right)^{1-\alpha_1}& \nu > \nu_o\\
\frac{2P_o\nu_o}{1-\alpha_2}\left(1-\frac{\nu}{\nu_o}\right)^{1-\alpha_2} + <Var(\nu_o
)> & \nu < \nu_o
\end{array}\right. \label{e:112}
\end{equation}
Equation (\ref{e:112}) may be effectively solved for $P(\nu)$ as long as 
$\nu P(\nu)$ is monotonically decreasing with $\nu$. X-ray data at hand seem to
be consistent with that.  We substitute observed X-ray variances into
Eq. (\ref{e:112}) and solve for parameters $\alpha$, $P_o$ and $\nu_o$.

\ \\
This paper has been processed by the authors using the Blackwell
Scientific Publications \LaTeX\  style file.


\bigskip

 


\begin{thebibliography}{}
\bibitem[]{} Abraham, R.G., McHardy, I.M., 1989, in 23rd ESLAB Symposium on
    Two topics in X-ray Astronomy: 2. AGN and the X-ray Background, 
    ESA-SP 292, page 865 
\bibitem[]{} Antonucci, R.R.J., 1993,  ARA\&A, 31, 473
\bibitem[]{} Begelman, M.C., McKee, C.F., Shields, G.A., 1983, ApJ, 271, 70
\bibitem[]{} Belloni, T., Hasinger, G., 1990, A\& A., 227, L33
\bibitem[]{} Clavel, J. et al., 1992, ApJ, 393, 113
\bibitem[]{} Collin-Souffrin, S., 1987, A\&A, 179, 60
\bibitem[]{} Collin-Souffrin, S., 1991, A\&A, 249, 344
\bibitem[]{} Cui, W., Zhang, S.N., Focke, W., Swank, J.H., 1997, ApJ, 484, 383
\bibitem[]{} Czerny, B., Lehto, H.J., 1997, MNRAS, 285, 365 
\bibitem[]{} Deeming, T.J., 1975, ApSS, 36, 137 and an errata.
\bibitem[]{} Done, C., Pounds, K.A., Nandra, K., Fabian, A.C., 1995, MNRAS, 
     275, 417 
\bibitem[]{} Gondek, D., Zdziarski, A.A., Johnson, W.N., George, I.M., 
    McNaron-Brown, K., Magdziarz, P., Smith, D., Gruber, D.E., 1996, MNRAS, 
    282, 646
\bibitem[]{} Green, A.R., McHardy, I.M. \& Lehto, H.J., 1993 MNRAS, 265, 664
\bibitem[]{} GreenhillL, L.J., Gwinn, C.R., Antonucci, R., Barvainis, R., 1996,
    ApJL, 472, L21
\bibitem[]{} Kaastra, J.S., Barr, P., 1989, A\&A, 226, 59 
\bibitem[]{} Kazanas, D., Hua, X.,Titarchuk, L.,  1997, ApJ, 480, 735
\bibitem[]{} Korista, K.T., et al., 1995, ApJS, 97, 285  
\bibitem[]{} K\"{o}nig, M. \& Timmer, J., 1997, A\&AS, 124, 589
\bibitem[]{} Kuraszkiewicz, J., Loska, Z., Czerny, B., 1997, Acta Astr.,
      47, 263
\bibitem[]{} Kurpiewski, A., Kuraszkiewicz, J., Czerny, B., 1997, MNRAS, 285,
      725
\bibitem[]{} Lawrence, A., Pounds, K.A., Watson, M.G., Elvis, M.S.,  
        1987, Nat., 325, 692 
\bibitem[]{} Loska, Z., Czerny, B., 1997, MNRAS, 284, 946
\bibitem[]{} Magdziarz, P., Blaes, O., Zdziarski, A.A., Johnson, W.N., 
    Smith, D.A., 1997, in Proceedings of the Fouth Compton Symp., eds.
    C.D. Dermer, J.K. Kurfess and M.S. Strickman (in press)
\bibitem[]{} Marshall, H.L., Fruscione, A., Carone, T.E., 1995, ApJ, 439, 90
\bibitem[]{} Marshall, H.L. et al. 1997, ApJ, 479, 222
\bibitem[]{} McHardy, I.M., Czerny, B., 1987, Nat., 325, 696
\bibitem[]{} McHardy, I.M., 1989,  in 23rd ESLAB Symposium on
    Two topics in X-ray Astronomy: 2. AGN and the X-ray Background, 
    ESA-SP 292, page 1111
\bibitem[]{} Mushotzky, R.F., Done, C., Pounds, K.A., 1993, ARA\&A, 31, 717
\bibitem[]{} Mushotzky, R.F., Fabian, A.C., Iwasawa, K., Kunieda, H., 
    Matsuoka, M., Nandra, K., Tanaka, Y., 1995, MNRAS, 272, L9
\bibitem[]{} Nandra, K., George, I.M., Mushotzky, R.F., Turner, T.J., 
    Yaqoob, T., 1997a, ApJ, 477, 602
\bibitem[]{} Nandra, K., Mushotzky, R.F., Yaqoob, T., George, I.M., Turner, 
    T.J., 1997b, MNRAS, 284, L7
\bibitem[]{} Nandra, K, Pounds, K.A., 1994, MNRAS, 268, 405
\bibitem[]{} Narayan, R., McClintock, J.E., Yi, I., 1996, ApJ, 431, 359
\bibitem[]{} Ostriker, E.C., McKee, C.F., Klein, R.I., 1991, ApJ, 377, 593
\bibitem[]{} Papadakis, I., Lawrence, A., 1993, Nat., 361, 233 
\bibitem[]{} Papadakis, I., Lawrence, A., 1995, MNRAS, 272, 161
\bibitem[]{} Papadakis, I. and M$\rm^{c}$Hardy, I.M., 1995,
        MNRAS,  273, 923-939
\bibitem[]{} Peterson, B.M. et al., 1991, ApJ, 368, 119
\bibitem[]{} Peterson, B.M. et al., 1992, ApJ, 392, 470
\bibitem[]{} Peterson, B.M. et al., 1994, ApJ, 425, 622
\bibitem[]{} Pounds, K.A., Nandra, K., Steward, G.C., George, I.M., Fabian, 
      A.C., 1990, Nat., 344, 132
\bibitem[]{} Rokaki, E., Collin-Souffrin, S., Magnan, C., 1993, A\&A, 272, 8
\bibitem[]{} R\' o\. za\' nska, A., Czerny, B., \. Zycki, P.T. Pojma\' nski, 
   G., 1998 (submitted to MNRAS) 
     
\bibitem[]{} Sergeev S.G., Pronik, V.I., Malkov, Yu.F., Chuvaev, K.K., 
       1997, A\&A, 320, 405
\bibitem[]{} Shakura, N.I., Sunyeav, A.R., 1973, A\&A, 24, 337
\bibitem[]{} Shakura, N.I., Sunyaev, R.A., 1976, MNRAS, 175, 613
\bibitem[]{} Siemiginowska, A., Czerny, B., Kostyunin, V, 1996, ApJ, 
       458, 491
\bibitem[]{} Silanp\" a\" a, A., et al. 1996, A\&A, 305, L17
\bibitem[]{} Tanaka, Y., et al., 1995, Nat., 375, 659
\bibitem[]{} Tsonis, A.A., Elsner, J.B., 1992, Nat., 358, 217
\bibitem[]{} Van der Klis, M., 1995, in X-ray Binaries, eds. W.H.G. Lewin,
    J. van Paradijs \& E.P.J. van den Heuvel, Cambridge University Press,
    p. 252
\bibitem[]{} Vaughan, B., Nowak, M., 1997, ApJ, 474, L43
\bibitem[]{} Weaver, K.A., Yaqoob,T., Mushotzky, R.F., Nousek, J., Hayashi, I.,
    Koyama, K., 1997, ApJ, 474, 675
\bibitem[]{} Witt, H.J., Czerny, B., \. Zycki, P.T., 1997, MNRAS, 286, 848
\bibitem[]{} Yaqoob, T., Edelson, R., Weaver, K., Warwick, R.S., Mushotzky, 
    R.F., Serlemitsos, P.J., Holt, S.S., 1995, ApJ, 453, L81
\bibitem[]{} Yaqoob, T., Serlemitsos, P.J., Turner, T.J., George, I.M., 
    Nandra, K., 1996, ApJ, 470, L27
\bibitem[]{} Yaqoob, T., McKernan, B., Ptak, A., Nandra, K., Serlemitsos,
    P.J., 1997, ApJL (in press)  
\bibitem[]{} Zdziarski, A.A., Johnson, W.N., Magdziarz, P., 1996, MNRAS, 
    283, 193
\bibitem[]{} Zdziarski, A.A., Johnson, W.N., Poutanen, J., Magdziarz, P., 
    Gierli\' nski, M., 1997, CAMK preprint 316 
\end{thebibliography}
\end{document}